\documentclass[prl,notitlepage,twocolumn]{revtex4}

\usepackage{amsmath}
\usepackage{amssymb}
\usepackage{bm}
\usepackage{graphicx}
\usepackage{times}
\usepackage{xcolor}
\usepackage{float}
\usepackage{gensymb}

\usepackage{pdfcomment}
\pdfcommentsetup{draft} 
\pdfcommentsetup{color={1.0 1.0 0.0}, open=true}

\usepackage{color}
\usepackage{ulem}

\begin{document}

\title{Transverse scattering and generalized Kerker effects in all-dielectric Mie-resonant meta-optics}

\author{Hadi K. Shamkhi$^1$}
\author{Kseniia V. Baryshnikova$^1$}
\author{Andrey Sayanskiy$^1$}
\author{Polina Kapitanova$^1$}
\author{Pavel D. Terekhov$^{1,2,3,4}$}
\author{Pavel Belov$^1$}
\author{Alina Karabchevsky$^{1,2,3,4}$}
\author{Andrey B. Evlyukhin$^{5}$}
\author{Yuri Kivshar$^{1,6}$}
\author{Alexander S. Shalin$^1$}

\affiliation{$^1$ITMO University, St. Petersburg 197101, Russia}
\affiliation{$^2$Electrooptics and Photonics Engineering Department, Ben-Gurion University, Beer-Sheva 8410501, Israel}
\affiliation{$^3$Ilse Katz Institute for Nanoscale Science $\&$ Technology, Ben-Gurion University, Beer-Sheva 8410501, Israel}
\affiliation{$^4$Center for Quantum Information Science and Technology, Ben-Gurion University, Beer-Sheva 8410501, Israel}
\affiliation{$^5$ Institute of Quantum Optics, Leibniz University,  Hannover 30167, Germany}
\affiliation{$^6$Nonlinear Physics Center, Australian National University, Canberra ACT 2601, Australia}

\begin{abstract}
All-dielectric resonant nanophotonics lies at the heart of modern optics and nanotechnology due to the unique possibilities to control scattering of light from high-index dielectric nanoparticles and metasurfaces. One of the important concepts of dielectric Mie-resonant nanophotonics is associated with the Kerker effect that drives the unidirectional scattering of light from nanoantennas and Huygens' metasurfaces. Here we suggest and demonstrate experimentally a novel effect manifested in {\it  the nearly complete simultaneous suppression} of both forward and backward scattered fields.  This effect is governed by the Fano interference between an electric dipole and off-resonant quadrupoles, providing necessary phases and amplitudes of the scattered fields to achieve the transverse scattering. We extend this concept to dielectric metasurfaces that demonstrate zero reflection with transverse scattering and strong field enhancement for resonant light filtering, nonlinear effects, and sensing.
\end{abstract}

\maketitle

{\it Introduction.} Light scattering by subwavelength particles is closely associated with the Mie resonances and optically-induced multipolar response~\cite{Kruk2017,kiv}. Co-existence and interplay of electric and magnetic dipolar modes make it possible to achieve either constructive or destructive interference leading to remarkable variety of scattering patterns of subwavelength dielectric particles~\cite{Terekhov2017,Liu2018}. In particular, strong forward-to-backward asymmetric scattering (often termed as {\it the Kerker effect}) is achieved as a result of interfering electric and magnetic dipole modes~\cite{Kerker1983a,Nieto-Vesperinas2011,Krasnok2011,Dubois2018,OptLett835} or quadrupole modes with appropriate phase relations~\cite{Liu:14,Pors2015,Liu2018}. The Kerker effect is also of  a great interest to the near-field directionality of Huygens' dipoles~\cite{PhysRevLett.120.117402}, and it drives the physics of highly efficient Mie-resonant passive and active dielectric metasurfaces and metadevices~\cite{Decker2015,Kruk2016,Genevet2017,active,Terekhov_B}.

Overlapping the electric and magnetic {\it multipolar resonances of higher orders} opens the way for novel strategies in the effective shaping of light beyond the conventional forward and backward directions~\cite{Liu2018}. For example, an isolated $V$-shaped plasmonic nanoantenna~\cite{Li2016} or  a nanoparticle trimer~\cite{Lu2015} were suggested to achieve the side-directed scattering by breaking the scatterer's symmetry. Moreover, the specific conditions for the simultaneous cancellation of both forward and backward scattering were obtained in the quasi-static approximation for radially anisotropic particles~\cite{Ni2013}. However, in order to satisfy the power conservation and suppress the forward scattering at the same time, the suggested particles should possess a gain.

Recently, the scattering of three-layer onion-like nanoparticles were optimized to achieve anisotropic elliptical side-scattering patterns and suppress both forward and backward scattering~\cite{Jeng2018}. However, the described scenario appears to be not generic, and the layered geometry looks too complicated and not feasible. 

\begin{figure}[h!]
\centering
\includegraphics[width=1\linewidth]{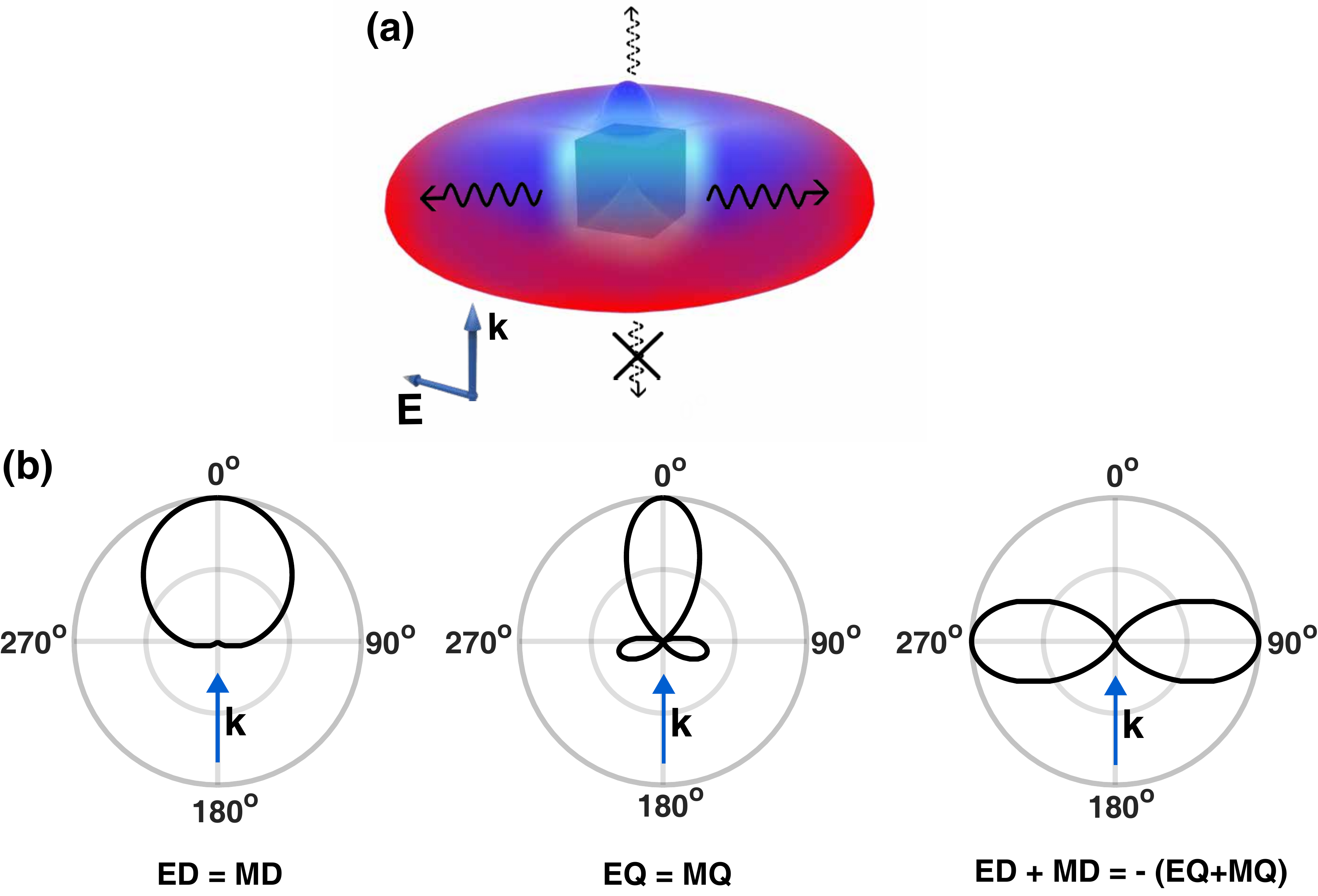}
\caption{(a) Transverse scattering pattern for a cubic nanoparticle. Small forward scattering corresponds to a contribution of higher-order multipoles in accordance with the optical theorem. (b) Concept of the formation of an ideal transverse scattering pattern. Electric dipole (ED) is in-phase with a magnetic dipole (MD), and an electric quadrupole (EQ) is in-phase with a magnetic quadrupole (MQ), whereas the dipoles are out-of-phase with the quadrupoles.}
\label{fig:figure1}
\end{figure}

Here we predict and demonstrate {\it a new general effect in high-index resonant meta-optics} characterized by the transverse isotropic scattering by subwavelength nanoparticles with the simultaneous suppression of both forward and backward scattered fields. We consider the overlapping of {\it four multipoles} (two dipoles and two quadrupoles, see Fig.~1) and reveal the physical mechanisms governing the transverse scattering, providing  the required phase and amplitude matching of the scattered fields. We carry out the proof-of-principle microwave experiment (being possible due to the scalability of the dielectric particles) and demonstrate a nearly perfect agreement with our analytical results based on the Mie theory~\cite{PENA20092348} and our full-scale numerical calculations~\cite{COMSOLMultiphysics}. Moreover, we design nearly invisible metasurfaces with zero reflection and near 100\% transmission governed by the simultaneous cancellation  of both forward and backward scattering in each meta-atom. The transverse scattering is accompanied by a strong near-field localization inside the particles, and it is conceptually similar to the optical anapoles~\cite{anapole,anapole2}. Such interference-driven effects are of a high demand for various applications including four-wave mixing~\cite{Maier2017} and enhanced second-harmonic generation~\cite{Kirill2018}.

{\it Concept.} We start with the Cartesian multipole decomposition of the field scattered by an arbitrary subwavelength particle. The surrounding medium is a free space with relative permittivity $\varepsilon =1$. The incident wave is assumed to be linearly polarized along the $x$-axis, $\mathbf{E}_{inc}=E_{0}e^{ikz}\mathbf{x}$, where $k=\vert\textbf{k}\vert$ is the wavenumber, and $\mathbf{x}$ is a unit vector along the $x$ axis. The scattered light is defined by a superposition of multipoles (up to the quadrupole terms)~\cite{jackson1998},
\begin{eqnarray}
\label{eq1}
\begin{gathered}
\textbf{E}_{sca}(\textbf{n})\cong \frac{k^{2} e^{i\textbf{k}\cdot\textbf{r}}}{4\pi r\varepsilon _{0}}\bigg(\lbrack 
\textbf{n}\times \lbrack \textbf{p}\times \textbf{n}\rbrack \rbrack
+\frac{1}{c}\lbrack \textbf{m}\times 
\textbf{n}\rbrack  \\ 
+\frac{ik}{6}\lbrack \textbf{n}\times \lbrack \textbf{n}\times 
(\hat{{Q}} \cdot \textbf{n}) \rbrack \rbrack +\frac{ik}{2c}\lbrack \textbf{n}\times 
(\hat{{M}}\cdot \textbf{n})\rbrack \bigg),
\end{gathered}
\end{eqnarray}
where $\mathbf{n}=\mathbf{r}/r$ is the unit vector directed from the particle's center towards an observation point; $c$ is the speed of light, $\mathbf{p}$($
\mathbf{m}$),~$\hat{Q}(\hat{M})$ are  the electric (magnetic) dipole and electric (magnetic) quadrupoles, respectively. Below, we consider {\it two different cases}: spherical particles described by the Mie theory, for which Eq.~(\ref{eq1}) can be simplified in terms of polarizabilities, and nonspherical particles described by the set of tensors from Eq.~(\ref{eq1}). For spherical particles, we can define te electric and magnetic dipolar ($\alpha_p$, $\alpha_m$) and quadrupolar ($\alpha_Q$, $\alpha_M$) polarizabilities. Utilizing the positive sign convention  $  \nabla \times \mathbf{E}_{inc}=i ck \mathbf{H}_{inc}$, we obtain~\cite{jackson1998,PhysRevB.79.235412}:
\begin{eqnarray}
\label{eq2}
\nonumber && \mathbf{p}=\alpha _{p}\mathbf{E}_{inc}; \mathbf{m}=\alpha _{m}\mathbf{H}_{inc}; \hat{Q}=\alpha 
_{Q}\frac{ \nabla \mathbf{E}_{inc}+( \nabla \mathbf{E}_{inc})^{T}}{2};\\ && 
\hat{M}=\alpha _{M}\frac{ \nabla \mathbf{H}_{inc}+( \nabla 
\mathbf{H}_{inc})^{T}}{2}.
\end{eqnarray}

Next, we derive the conditions for the simultaneous suppression of the forward and backward scattering. To do so, we define the angular distribution of the differential scattering cross-section by a spherical particle, up to the dipole and quadrupole terms~\cite{jackson1998}:

\begin{align}
\label{eq3}
\frac{d\sigma}{d\Omega }(\theta )\cong
\bigg\vert \frac{\alpha_{p}}{\varepsilon _{0}}+\alpha _{m}\cos \theta+\frac{k^{2}}{4}\bigg(\frac{\alpha _{Q}}{3\varepsilon 
_{0}}\cos \theta  
+\alpha _{M}\cos 2\theta\bigg)\bigg\vert^2,
\end{align}
where $\theta$ is the polar angle, and the power distribution is  considered to be symmetric in the azimuthal angle plane. Upon inspection of Eq.~(\ref{eq3}), find that the coherent scattering from the dipoles is directed mainly in the forward or backward  half-spaces depending on whether the term  $\Re\lbrace \alpha _{p}\alpha_{m}^{*}\rbrace $ is positive or negative, respectively~\cite{Nieto-Vesperinas2011}. Therefore, to obtain the total suppression of the backward scattering $\vert \theta \vert \ge 90^{\circ}$ (see Fig.~1), the dipoles have to be in-phase and have the same polarizabilities. For suppression of the forward scattering ($\vert \theta 
\vert \le 90^{\circ}$), the dipoles should also have the same 
polarizabilities, however the phase should be shifted by $\pi$ (the out-of-phase case). These cases are known as the Kerker and anti-Kerker conditions~\cite{Nieto-Vesperinas2011}. The former can be formulated as:
\begin{eqnarray}
\label{eq4}
\vert \alpha _{p}\vert =\vert\varepsilon _{0} \alpha _{m}\vert,\;\;\; \arg(\alpha _{p})=\arg(\alpha _{m})+2\pi n,
\end{eqnarray}
where $n$ is an integer. For quadrupole only coherent scattering, similar to the case of dipoles, the term  $\Re\lbrace\alpha _{Q}\alpha _{M}^{*}\rbrace$  governs asymmetry of the scattering, but with side lobes as shown in Fig.~1, and they can be called "quadrupolar Kerker conditions"~\cite{Pors2015,Liu2018},
\begin{equation}
\label{eq5}
\vert \alpha _{Q}\vert =\vert 3\varepsilon _{0}\alpha _{M}\vert, \; \arg(\alpha _{Q})=\arg(\alpha _{M})+2\pi n,
\end{equation}
 where $n$ is an integer. At last, the dipole-quadrupole overlapping is defined by the coupling term $\Re\lbrace(\alpha _{p}+\varepsilon _{0}\alpha _{m})(\alpha _{Q}+3\varepsilon _{0}\alpha _{M})^{*}\rbrace$. In a particular scattering scenario, if the conditions (\ref{eq4}) and (\ref{eq5}) are satisfied simultaneously, two scattering patterns can be realized depending on the phase relation between the dipoles and quadrupoles, while considering their equal or comparable amplitudes,
\begin{figure}[!b]
    \centering
    \includegraphics[width=\linewidth]{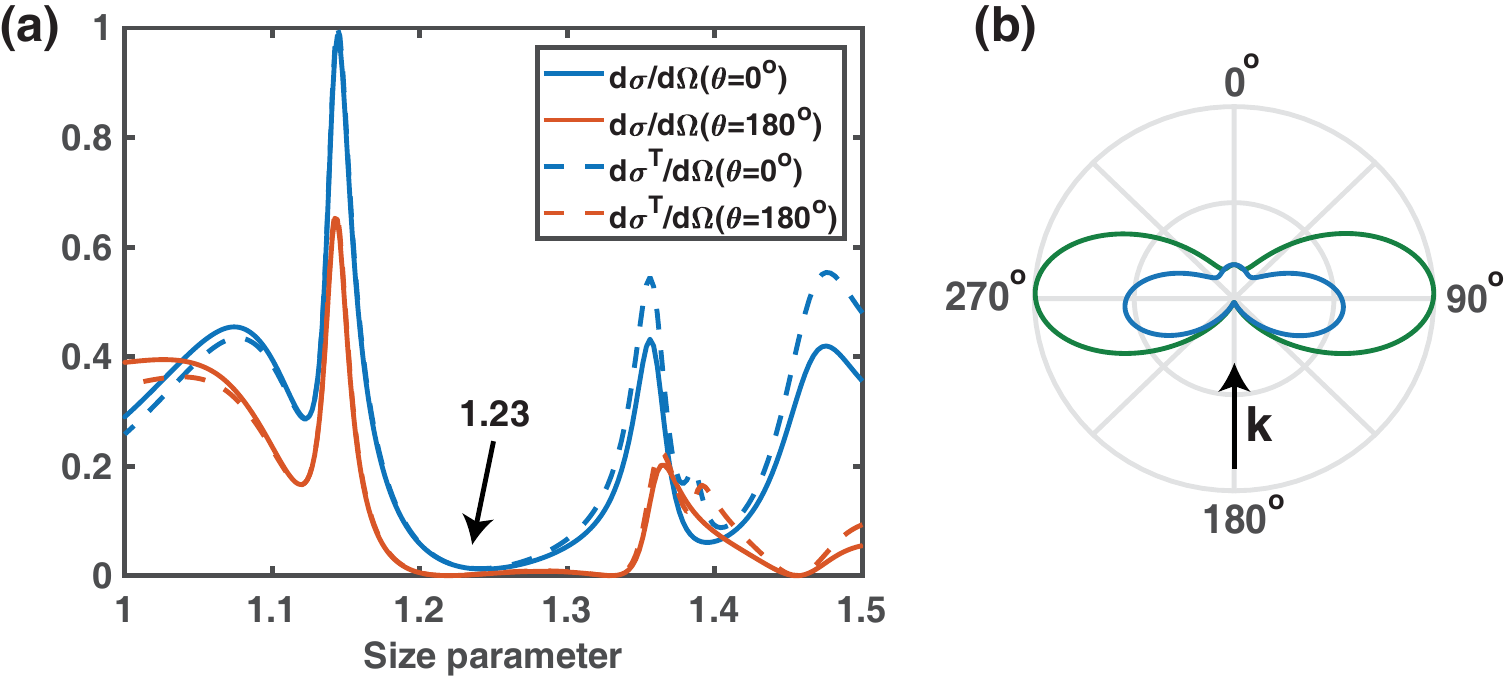}
    \caption{(a) Normalized differential scattering cross-sections calculated with the Mie theory for a silicon sphere of 120~nm radius in free space as functions of the size parameter $= k\cdot radius$. The solid lines are the calculations for dipole-quadrupolar sphere [${d\sigma }/{d\Omega }$, Eq.~(\ref{eq3})] while the dashed lines describe the total set of multipoles [${d\sigma^T }/{d\Omega }$, Eq.~(\ref{eq7})].  In the legend, $\theta$ = 0$^o$ refers the forward direction, and $\theta$ = 180$^o$-- to the backward direction. (b) Transverse scattering patterns at $609 $~nm wavelength with permittivity $15.44+0.196i$ ($1.23$ size parameter)~\cite{Aspnes1983}: the green (blue) line corresponds to the plane of the incident electric (magnetic) field polarization.}
    \label{fig:figure2}
  \end{figure}
\begin{eqnarray}
\label{eq6}
&\vert\alpha _{p}+\varepsilon _{0}\alpha _{m}\vert  =
 \frac{k^{2}}{12}\vert\alpha _{Q} 
+3\varepsilon
_{0}\alpha _{M}\vert ,\nonumber\\
 &\arg \left(\alpha _{p}+\varepsilon _{0}\alpha _{m} \right) =\pm 
\arg \left(\alpha _{Q} 
+3\varepsilon
_{0}\alpha _{M} \right).
\end{eqnarray}
The first condition (with 'plus') corresponds to the constructive interference of the dipoles and quadrupoles with enhanced directivity in the forward direction  (see, e.g., Ref.~\cite{Pors2015}). The second scattering regime described by Eq.~(\ref{eq6}) (with 'minus') corresponds to the case when destructive interference between the combined coherent dipoles and combined coherent quadrupoles nearly suppresses both forward and backward scattering. This kind of interaction leads to the formation of {\it the transverse scattering pattern}, as shown in Fig.~1. The side lobes of the quadrupoles form a scattering pattern in the lateral plane. However, for realistic subwavelength particles, it is impossible to achieve full cancellation because of the optical theorem that links the total extinction cross-section of a particle to the forward-scattered fields. Therefore, although the forward scattering cannot be completely eliminated, it can be suppressed significantly if the conditions (\ref{eq4})-(\ref{eq6}) are satisfied. Figure~1(b) presents this concept schematically. 

Moreover, we can generalize the conditions (\ref{eq4})-(\ref{eq6}) to the whole set of multipoles, referring to them as {\it the generalized Kerker and anti-Kerker conditions}, respectively~\cite{Liu2018,Liu:14}
\begin{equation}
\label{eq7}
\frac{d\sigma ^{T}}{d\Omega }(\theta =180^{o})=0, \;\;\;\;
 \frac{d\sigma ^{T}}{d\Omega }(\theta =0^{0})\cong 0.
\end{equation}
The normalized differential scattering cross-sections for a spherical Si particle of 120 nm radius are shown in Fig.~\ref{fig:figure2},  where we employ the Mie theory and experimental data~\cite{Aspnes1983} to solve Eq.~(\ref{eq3}) for the multipole contributions up to the magnetic quadrupole; and compare it with a sum of all multipole contributions.  The conditions (\ref{eq4})-(\ref{eq7}) are almost satisfied at the wavelength of 609~nm (size parameter 1.23), when the backward intensity vanishes and the forward scattering intensity reaches its minimum.  The result in Fig.~\ref{fig:figure2} proves that the contribution of the higher-order multipoles  to the scattered field is negligible and, thus, the dipole-quadrupole approximation remains valid.%
\begin{figure}[tb]
    \centering
    \includegraphics[width=\linewidth]{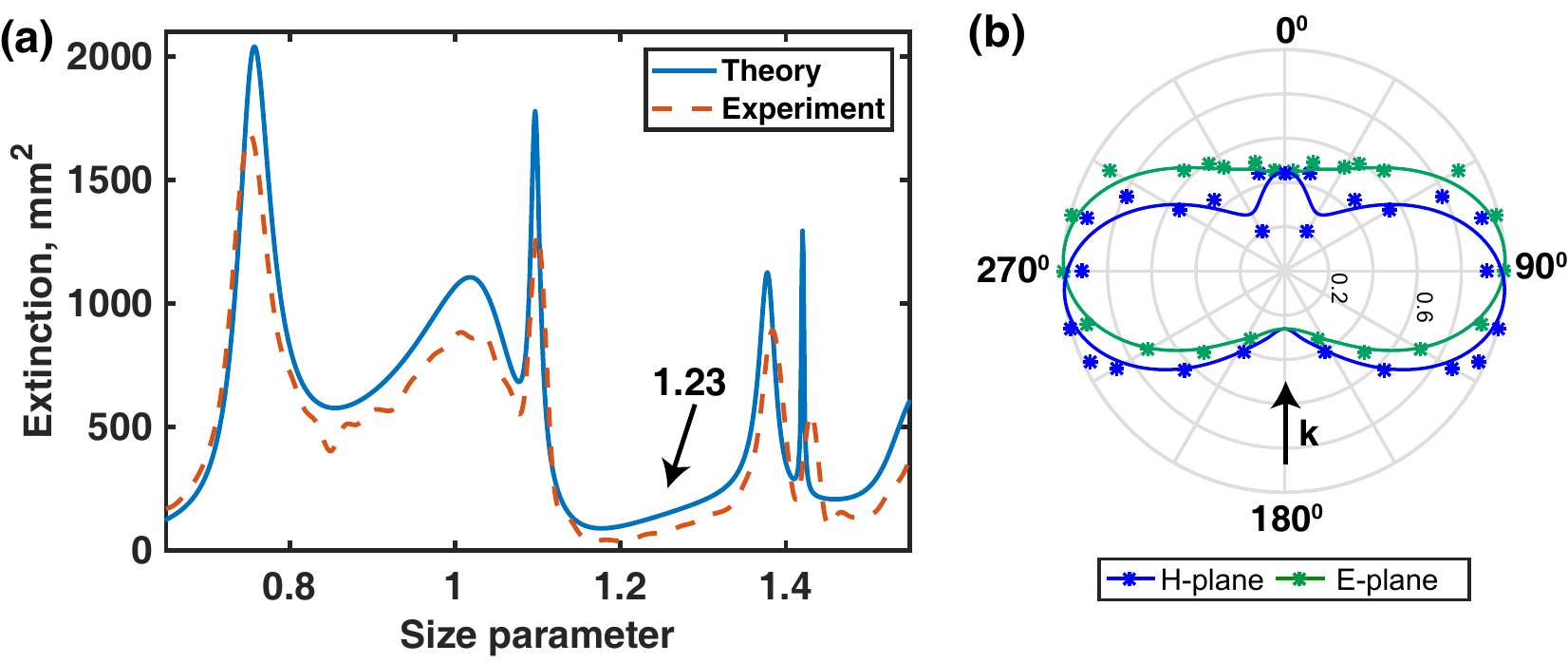}
    \caption{(a) Comparison of the extinction cross-section calculated with the Mie theory and measured experimentally. (b) Scattering patterns at the frequency $f$ = 7.85 GHz (1.23 size parameter), where the green (blue) curve corresponds to the plane of the incident electric (magnetic) fields polarization. Results are obtained for the ceramic sphere with the relative dielectric permittivity 16+0.001i and radius 7.5 mm, respectively.}
    \label{fig:figure3}
  \end{figure}
  
{\it Experimental results.} Based on Maxwell's equations, we verify our concept by measuring the extinction cross-section and scattering patterns of a spherical particle at the microwave frequency range. To mimic the scattering properties of silicon nanoparticles at microwaves, we use MgO-TiO$_2$ ceramic spheres characterized by the dielectric constant $\varepsilon=16$  and dielectric loss factor of about 0.00112, measured in the range 9-12~GHz~\cite{Kanareykin2003}. A spherical particle with the radius $a$=7.5~mm is located in a microwave chamber for the measurements in the frequency range 4-10~GHz. To approximate a plane wave excitation, we use a rectangular horn antenna (TRIM 0.75-18 GHz; DR) connected to the transmitting port of a vector network analyzer (Agilent E8362C). The ceramic sphere is located in the far-field of the antenna, at the distance of approximately 2.5 m, and the second horn antenna (TRIM 0.75 - 18~GHz; DR) is used as a receiver to observe the transmitted signal. Forward scattering is calculated from the transmission coefficient. The total extinction cross-section is extracted from the measured complex magnitude of the forward scattered signal by means of the optical theorem~\cite{Newton1976}. The measured extinction is compared with the theoretically obtained results (Mie theory) in Fig.~3(a). To measure the two-dimensional scattering diagram, we change slightly the experimental setup. The transmitting antenna and the position of the ceramic sphere remain fixed, whereas the receiving antenna is moved  around the spherical particle in the $x-z$ plane. The scattering cross-section patterns for both theoretical (Mie theory) and measured data in the $x-z$ plane are shown in Fig.~3(b). We observe the lateral scattering to occur for the broadband off-resonance region from $f =7.6$~GHz to $f =7.9$~GHz. 

{\it Nonspherical nanoparticles.} Next, we extend our results to a more general case of nonspherical particles by employing the  Cartesian multipole moments~\cite{raab2005multipole,Vrejoiu2002}. After applying the procedure analogous to that described above, the conditions for the transverse scattering under $x$-polarized plane wave illumination can be written in the following form~\cite{suppl}
\begin{eqnarray}
\label{eq8}
cp_{x}/m_{y}=1; cQ_{xz}/ 3M_{yz}=1; 2icp_{x}/kM_{yz}\cong-1.
\end{eqnarray}
\begin{figure}[!b]
\centering
\includegraphics[width=\linewidth]{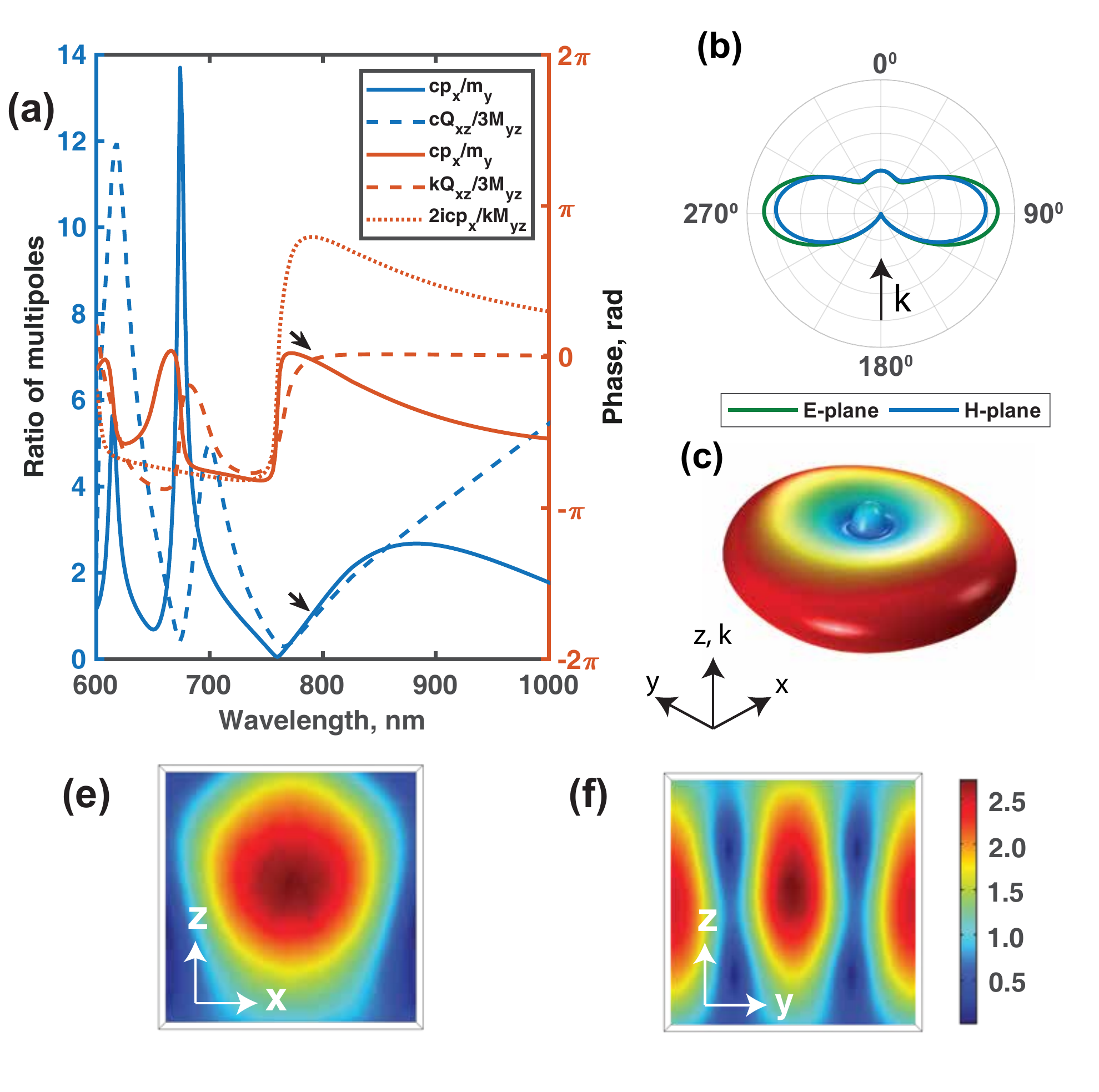}
\caption{(a) Amplitudes and phases of the multipole ratios from Eq.~(\ref{eq8}) for a cubic Si nanoparticle with the edge of 250 nm. The arrows show the wavelength corresponding to the transverse scattering  $\lambda$=788 nm. (b,c) Two- and three-dimensional scattering patterns at $\lambda$=788 nm. (e,f) Calculated electric field inside the nanoparticle with the transverse scattering pattern, $\lambda$ = 788 nm in the $xy$ and $xz$ planes, respectively. All calculations were conducted by COMSOL Multiphysics~\cite{COMSOLMultiphysics} and the silicon permittivity from \cite{Aspnes1983}.}
\label{fig:figure4}
\end{figure}
We notice that Ref.~\cite{Terekhov2017} a lateral-like
scattering was revealed as an “incidental finding” in the study of scattering of Si nanoparticles. However, that paper provided neither systematic study nor detailed physical explanation. Figure 4(a) shows the calculated absolute values of the ratios from Eq.~(\ref{eq8}) and their phases being the phase differences between the involved multipoles. In the wavelength range $700 - 820 $~nm, the electric dipole has sharp Fano profile \cite{Limonov2017} while the magnetic dipole increases monotonically, therefore the dipoles have the form of two crossing lines. On the other han, for the wavelength range $\lambda\geq 780 $ nm, the quadrupoles ratio shows linear  behaviour with constant zero-phase difference. This behavior is actually associated with the long wavelength regime of the quadrupoles. At the particular wavelength  $\lambda= 788 $~nm which is a shared point for dipoles ratio Fano profile and quadrupoles long-wavelength region   (black arrows), one can see that the dipoles are in-phase (solid red line) with  the nearly equal amplitudes (solid blue line) indicating the Kerker  effect. At the same time, the quadrupoles, have comparable values:  $cQ_{xz}/3M_{yz}=0.94$ (dashed blue curve), and are in phase (dashed red curve), which explains the generalized Kerker effect working on the assumption that higher multipoles are negligible. The phase difference between the coherent dipoles and quadrupoles (dotted red line) is about $0.75\pi$. The corresponding scattering patterns are shown  in Figs.~\ref{fig:figure4}(b,c). Hence, we observe almost complete scattering suppression in the forward and backward directions. However, the suppression of the forward scattering is not complete owing to the optical theorem.  As a result, we now uncover the physical requirements for the transverse scattering along with the mathematical conditions (8): the Fano interference of a dipole mode and off-resonant, long-wavelength quadrupoles. These mechanisms provide the necessary phases and amplitudes of the scattered fields in order to obtain the required optical signature. In Supplementary Information~\cite{suppl}, we develop this approach further and take into account the presence of a substrate. 

In Figs.~\ref{fig:figure4}(e,f), we demonstrate the strong electric field concentration in both the $y-z$ and $x-z$ planes inside the nanoparticle at the wavelength of the transverse scattering. The observed near-field enhancement is accompanied by the strong scattering suppression resembling the case of the anapole mode~\cite{anapole} where the dipole radiation is almost cancelled by the field of the electric toroidal moment. This effect of the localization of light could be of a great interest for nonlinear nanophotonics.  

{\it Metasurfaces}. Finally, we study a novel type of metasurfaces composed of cubic nanoparticles which allow satisfying the condition (\ref{eq8}) for the transverse scattering in the visible frequency range. Figure 5(a) shows schematically a square lattice of identical cubic  nanoparticles illuminated with a normally incident wave. In the far-field region, the reflection is determined by the effective polarizabilities and multipole moments of a central nanoparticle~\cite{Evlyukhin2010} taking into account the interaction with all other nanoparticles of the metasurface. However, this multipole description can be applied  independently to the metasurface with varying lattice spacing~\cite{Evlyukhin2010}, which suggest that the obtained conditions (\ref{eq8}) hold. 

\begin{figure}[tb]
    \centering
    \includegraphics[width=0.47\textwidth]{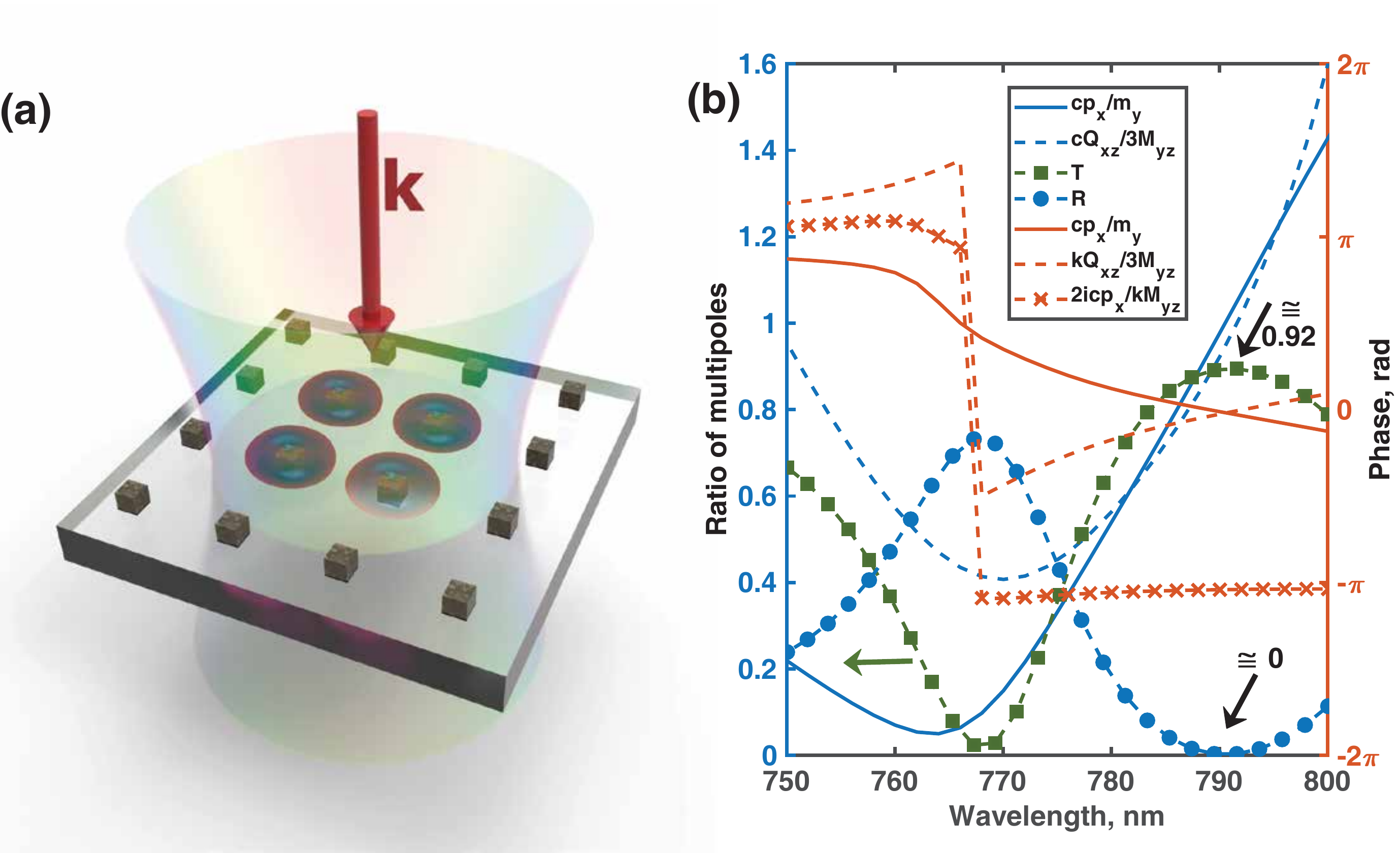}
    \caption{(a) Schematics of a metasurface being a lattice of cubic nanoparticles with strong transverse scattering. The substrate and the Gaussian beam are shown for better visibility; numerically we consider the metasurface in free-space irradiated with a plane wave. (b) Amplitudes  (blue) and phases (red) of the multipole ratios defined by Eq.~(\ref{eq8}) for Si cube with the edge of 250 nm and the period $p$ = 400 nm, and the corresponding reflection (R) and transmission (T) coefficients. Note, the reflection and transmission coefficients are plotted in accordance with the left $y$ axis.}
    \label{fig:figure5}
  \end{figure}
We perform the full-wave numerical simulations of metasurfaces with the lattice spacing $400 $~nm. Figure \ref{fig:figure5}(b) shows the ratios of multipoles (\ref{eq8}) in the case of a metasurface, similar to those shown in Fig.~\ref{fig:figure4} for an isolated particle. The polarizabilities are affected by coupling, but its multipole moments within the array still satisfy the transverse scattering conditions (\ref{eq8}). In the wavelength range of $787-793$~nm the coherent dipoles are nearly in-phase (solid red curve), and the quadrupoles are also in-phase (dashed red curve), but they are in out-of-phase with each other (red curve marked with "x"). With the close to unity amplitude ratios (solid and dashed blue curves), the metasurface demonstrates a nearly perfect simultaneous suppression of the forward and backward scattering, similar to an isolated nanoparticle (see Fig.~\ref{fig:figure4}).  Therefore, in this wavelength range, the metasurface reflection vanishes (blue dashed curve labeled with circles), despite the fact that light interacts with the structure providing strong near fields [see Figs. \ref{fig:figure4}(e,f)], and the transmission (green dashed curve labeled with cubes) is near-unity and 100 \% transmission can be obtained for the loss-less case. In contrast to the well-known Huygens' metasurfaces~\cite{Decker2015},  the novel metasurfaces introduced here  suppress almost completely scattering in both forward and backward directions being nearly invisible. For more information about the lattice spacing and a substrate impacts on the invisibility effect, see comment
\footnote{Increasing the lattice spacing leads to a broadening of the invisibility region. The inter-particle coupling becomes stronger and, as a result, the multipole resonances experience an incoherent shift of the spectrum. On the other hand, a substrate introduced to the metasurface will only shift the near-zero region to shorter wavelengths.} and the Supplementary Information~\cite{suppl}. 

{\it Conclusions}. We have demonstrated a novel effect of the transverse scattering of light by Mie-resonant subwavelength particles with the simultaneous suppression of both forward and backward scattering. This generalized Kerker effect occurs  when in-phase electric and magnetic dipoles become out of phase with the corresponding quadrupoles. We have obtained the general conditions for the simultaneous suppression of scattering in both forward and backward directions, and have generalized these conditions to nonspherial particles. We have revealed the crucial role played by the electric-dipole Fano interference with the off-resonant quadrupoles for achieving the transverse scattering, and we have verified our concept in a proof-of-principle microwave experiments, with a good agreement with the analytical and numerical results. Finally, we have studied the  metasurfaces composed of the nanoparticles radiating with the transverse scattering patterns and have demonstrated their periodicity-dependent zero reflection. In a sharp contrast to Huygens' metasurfaces, these novel metasurfaces scatter neither forward nor backward, being almost invisible. The predicted effects can be utilized for an efficient beam control,  strong field enhancement required for nonlinear interaction, and highly efficient sensing. 
 
\begin{acknowledgments}
The authors acknowledge a financial support from the Russian Foundation for Basic Research (grants 18-02-00414 and 18-52-00005); the Ministry of Education and Science of the Russian Federation (grant 3.4982.2017/6.7); the Israeli Ministry of Trade and Labor-Kamin Program (grant. No. 62045), and the Strategic Fund of the Australian National University. Experimental characterization of the high-index structures as well as investigation of anapole states were financially supported by Russian Science Foundation (grant 17-19-01731 and grant 17-72-10230, correspondingly). YK thanks Wei Liu for useful discussions and suggestions.
\end{acknowledgments}

\end{document}